\documentclass[10pt,journal]{IEEEtran}
\usepackage[left=0.6in,right=0.7in,top=0.5in,bottom=0.6in]{geometry}

\usepackage{latexsym}
\usepackage{graphicx}
\usepackage{epsfig}
\usepackage{subfigure}
\usepackage{array}
\usepackage{amsmath}
\usepackage{amssymb}
\usepackage{color, soul}
\usepackage{amsthm}
\usepackage{enumerate}
\usepackage{algpseudocode}
\usepackage{algorithm}
\usepackage{stfloats}

\newcommand{\bR}{\textbf R}

\newcommand{\bW}{\textbf W}
\newcommand{\bH}{\textbf H}
\newcommand{\bM}{\textbf M}

\newcommand{\bI}{\textbf I}

\newcommand{\bQ}{\textbf Q}
\newcommand{\bZ}{\textbf Z}
\newcommand{\bT}{\textbf T}
\newcommand{\bG}{\textbf G}
\newcommand{\bV}{\textbf V}
\newcommand{\bP}{\textbf P}
\newcommand{\bN}{\textbf N}
\newcommand{\bX}{\textbf X}

\DeclareMathOperator*{\argmax}{arg\,max}

\theoremstyle{plain} 

\newtheorem{prop}{Proposition}
\theoremstyle{definition}

\def\bal#1\eal{\begin{align}#1\end{align}}

\begin{document}

\title{Comments on "Precoding and Artificial Noise Design for Cognitive MIMOME Wiretap Channels"}

\author{Mahdi Khojastehnia, Sergey Loyka, \IEEEmembership{Senior Member, IEEE}

\vspace*{-1\baselineskip}

\thanks{M. Khojastehnia and S. Loyka are with the School of Electrical Engineering and Computer Science, University of Ottawa, Canada, e-mail: sergey.loyka@uottawa.ca}
}

\maketitle

\vspace*{-1\baselineskip}

\begin{abstract}
Several gaps and errors in [1] are identified and corrected. While accommodating these corrections, a rigours proof is given that the successive convex approximation algorithm in [1] for secrecy rate maximization (SRM) does generate an increasing and bounded sequence of true secrecy rates and hence converges. It is further shown that its convergence point is a KKT point of the original SRM problem and, if the original problem is convex, this convergence point is globally-optimal, which is not necessarily the case in general. An interlacing property of the sequences of the true and approximate secrecy rates is established.
\end{abstract}



\section{Introduction}

Secrecy rate maximization (SRM) in cognitive MIMOME wiretap Gaussian channels under artificial noise (AN) transmission was considered in \cite{Fang-16}. Since the underlying optimization problem is not convex and hence difficult to solve, a successive convex approximation (SCA) algorithm was proposed. In this note, we identify and correct the following major gaps and errors in \cite{Fang-16}:

1. Lemmas 1 and 2 in \cite{Fang-16} are based on the false assumption of strict convexity of problems (P3) and (P4): these problems are never strictly-convex since their objective is linear in $t$\footnote{Recall that a linear function is not strictly concave, see e.g. \cite[Ch. 3]{Boyd-04}.}, see \cite[eq.(15), eq.(20)]{Fang-16}, and hence is not strictly concave. This also applies to problem (P7).

2. It is claimed, without proof, that the key inequality (17) in \cite{Fang-16} holds due to strict convexity of (P3), which is incorrect.

3. While Algorithm 1 in \cite{Fang-16} computes the true secrecy rate $C_s(\bQ^v,\bW^v)$, its (incorrect) proof of convergence in Lemma 1 is using its concave approximation $R(\bQ^{v},\bW^{v})$. However, the convergence of the latter does not imply the convergence of the former. This also applies to Algorithm 2.

4. The notations $R(\bQ,\bW)$ and $R(\bQ, \bW|\bQ^v,\bW^v)$ in \cite[eq. (16), (17)]{Fang-16} and elsewhere are incorrect: since $R$ is the objective of (P3), see \cite[eq.(15)]{Fang-16}, it depends on $\bQ$ and $t$ only and, hence, should be $R(\bQ,t)$. Eq. (16)-(19) in \cite{Fang-16} should be modified accordingly\footnote{While $\bW$ may influence $\bQ$ indirectly, via the constraint in [1, eq. (15)], this happens only if this constraint is active and not otherwise. Furthermore, in the KKT conditions, the objective function and constraints are treated in different ways and, when gradients are computed, only explicit arguments are taken into account but not indirect relationships via constraints.}. Likewise, $\overline{\varphi}_k(\bQ,\bW)$ in (P3) and elsewhere should be $\overline{\varphi}_k(\bQ,\bW|\bQ^v,\bW^v)$ since it depends on its expansion point $(\bQ^v,\bW^v)$, see \cite[eq. (13)]{Fang-16}. The same applies to $\psi(\bQ,\bW)$ in \cite[eq. (21)]{Fang-16}. $\bT_k$ and $\bR_k$ in \cite[eq. (14)]{Fang-16} and elsewhere should be modified to $\bT_k(\bQ^v,\bW^v)$ and $\bR_k(\bQ^v,\bW^v)$ due to the same reason. This is important as the analysis and proof of convergence depend significantly on these missing arguments, which is a possible source of the errors in \cite{Fang-16}.

5. The feasible set in \cite[eq. (16)]{Fang-16} is not given - this is important as it changes from step to step, according to \cite[eq. (15)]{Fang-16}, where the expansion point $(\bQ^v,\bW^v)$ of $\overline{\varphi}_k(\bQ,\bW|\bQ^v,\bW^v)$ is updated at each step $v$ (note that, according to Comment 4 above, we are using here the correct notation). The optimization variables of this problem should include $t$ as well. A correct statement of this problem is given in \eqref{eq.Q^v,W^v,t^v} of this paper.

6. A termination criterion for Algorithms 1 and 2 is missing. Without it, a proof of their convergence is illusive: to prove the convergence of an algorithm, one has to demonstrate that its termination criterion is eventually satisfied (see e.g. the proofs of convergence for the algorithms in \cite{Boyd-04}).

These gaps and errors are corrected below. Additionally, the following novel contributions are provided:

7. In Proposition 2, we show rigorously that Algorithm 1 in \cite{Fang-16} does converge. This proof is significantly different from the unproved claims of Lemma 1 of \cite{Fang-16}. Additionally, this proposition also shows that both the true secrecy rate $C_s(\bQ^v,\bW^v)$ and its concave approximation $R(\bQ^v,t^v)$, are increasing bounded sequences and hence converge, and they satisfy an interlacing property.

8. Proposition 3 shows that a convergence point of $(\bQ^v,\bW^v)$ is a stationary (KKT) point of problem (P2) and hence of the original problem (P1) in in \cite{Fang-16}. Under certain condition, Proposition 4 shows that this point is globally optimum for (P1) and (P2) in \cite{Fang-16} (this is the only case known to us when Algorithm 1 is guaranteed to solve (P2) and hence (P1)).

Unless stated otherwise, we use the same channel model, assumptions and notations as in \cite{Fang-16}.

\section{Secrecy Rate Maximization in \cite{Fang-16}}

For completeness, we state below the optimization problems of \cite{Fang-16}. The SRM problem (P1) in \cite{Fang-16} can be expressed as follows:
\bal
\label{eq.(P1)}
(P1)\quad \ C_s^*=\max_{\bZ \in S_{Z}} C_s(\bZ)
\eal
where $C_s^*$ is the maximum achievable secrecy rate under AN transmission, $C_s(\bZ)$ is an achievable secrecy rate for a given $\bZ=(\bQ,\bW)$ (aggregate matrix variable), $\bQ$ and $\bW$ are the transmitted signal and artificial noise covariance matrices,
\bal
C_s(\bZ)=C_0(\bQ)-\max_{k \in \mathcal{K}}C_k(\bZ)
\eal
where $C_0(\bQ)=\log|\bI+\bH_0\bQ\bH_0^H|$ is an achievable rate of the Alice-Bob link, $C_k(\bZ)= \varphi_k(\bZ)-\phi_k(\bW)$ is $k$-th eavesdropper rate,
\bal
\label{eq.varphi_k}
\varphi_k(\bZ)&= \log|\bI+\bH_k\bQ\bH_k^H+\bP_k\bW\bP_k^H|\\
\label{eq.phi_k}
 \phi_k(\bW)&=\log|\bI+\bP_k\bW\bP_k^H|
\eal
$\bH_0, \bH_k$ are the Bob's and eavesdropper's channels, $\bP_k$ is the projected eavesdropper channel (on the null space of $\bH_0$). The feasible set $S_{Z}$ is as follows:
\bal\notag
\label{eq.S_Q,W}
S_{Z}=\big\{&\bZ: tr(\bG\bQ\bG^H+\bG\bV\bW\bV^H\bG^H) \le \Gamma,\\ &tr(\bQ+\bW) \le P, \ \bQ \succeq 0,\ \bW \succeq 0\}
\eal
where $\Gamma$ and $P$ represent the interference and transmit power constraints, $\bG$ is the primary user (PU) channel matrix and $\bV$ is the projection matrix on the null space of $\bH_0$. Note that $C_s^*= C_s(\bZ^*)$, where $\bZ^*=(\bQ^*, \bW^*)$ denotes optimal covariance matrices of signal and artificial noise.

To facilitate the algorithm design and analysis, problem (P1) was further transformed into the equivalent problem (P2) in \cite[eq. (9)]{Fang-16}:
\bal
\label{eq.P2}
(P2)\quad \ \max_{(\bZ,t) \in S_{Z,t}} R(\bQ,t)
\eal
where $R(\bQ,t)=C_0(\bQ)-t$, $t$ is a slack variable. The feasible set $S_{Z,t}$ is as follows:
\bal
\label{eq.S_Q,W,t}
S_{Z,t}&=\big\{(\bZ,t):\ \bZ \in S_z,\ C_k(\bZ) \le t \ \forall k \in \mathcal{K}\}
\eal
where, compared to \cite{Fang-16},  we eliminated the constrain $t\ge 0$\footnote{Unlike \cite{Fang-16}, we are using here $t \ge 0$ rather than $t>0$ since $t=0$ is also possible at optimal point, e.g. when $\bH_k=0$ for all $k$.} since it is redundant: $t\ge C_k(\bZ) \ge 0$.
Since $C_k(\bZ)$ are not convex, (P2) is not a convex problem and, hence, is difficult to solve \cite{Boyd-04}. It was further approximated by the following convex problem (P3) \cite[eq. (15)]{Fang-16}:
 \bal
\label{eq.P3}
(P3)\ \  \max_{(\bZ,t) \in S_{Z,t}(\bZ^v)} R(\bQ,t)
\eal
where the feasible set $S_{Z,t}(\bZ^v)$ is\footnote{We omitted the redundant constraint $t\ge 0$, since $t\ge C_{lk}(\bZ|\widetilde{\bZ})\ge C_k(\bZ)\ge 0$ (see Proposition 1 below).}
\bal
\label{eq.S_Q,W,t|Q^v,W^v}
S_{Z,t}(\bZ^v)=\big\{(\bZ,t):\ \bZ \in S_z,\ C_{lk}(\bZ|\bZ^v) \le t \ \forall k \in \mathcal{K} \}
\eal
where $C_{lk}(\bZ|\bZ^v)$ is a convex approximation of $C_k(\bZ)$ at $\bZ^v$:
\bal
C_{lk}(\bZ|\bZ^v)= \overline\varphi_k(\bZ|\bZ^v)-\phi_k(\bW)
\eal
where $\overline\varphi_k(\bZ|\bZ^v)$ is the linear approximation of $\varphi_k(\bZ)$ around $\bZ^v$:
\bal\notag
\overline\varphi_k(\bZ|\bZ^v)=\varphi_k(\bZ^v) &+ tr(\bT_k(\bZ^v)(\bQ-\bQ^v))\\
 &+ tr(\bR_k(\bZ^v)(\bW-\bW^v))
\eal
where $\bT_k(\bZ^v)$ and $\bR_k(\bZ^v)$ are the derivatives of $\varphi_k(\bZ)$  at $\bZ^v$ with respect to $\bQ$ and $\bW$ \cite[eq. (14)]{Fang-16}:
\bal\notag
\label{eq.TR} \notag
\bT_k(\bZ^v)&=\bH_k^H(\bI+\bH_k\bQ^v\bH_k^H+\bP_k\bW^v\bP_k^H)^{-1}\bH_k \\
\bR_k(\bZ^v)&=\bP_k^H(\bI+\bH_k\bQ^v\bH_k^H+\bP_k\bW^v\bP_k^H)^{-1}\bP_k
\eal

Finally, Algorithm 1 in \cite{Fang-16} computes iteratively the approximate optimal point $(\bZ^{v+1}, t^{v+1})$  at $(v+1)$-th iteration using the previous step approximation $\bZ^v$ as follows:
\bal
\label{eq.Q^v,W^v,t^v}
\mbox{(P3a)}\quad (\bZ^{v+1},t^{v+1}) =\argmax_{(\bZ,t) \in S_{Z,t}(\bZ^{v})} R(\bQ,t)
\eal
where, compared to \cite[eq.(16)]{Fang-16}, we have corrected the notations and arguments following the comments in the Introduction and the above discussion. Note that the feasible set $S_{Z,t}(\bZ^{v})$ here is not the same at each iteration but rather depends on $\bZ^{v}$ from the previous iteration, due to the constraint $C_{lk}(\bZ|\bZ^{v}) \le t$ in \eqref{eq.S_Q,W,t|Q^v,W^v} while the objective $R(\bQ,t)$ is the same at each iteration, i.e. the same objective is optimized over iteratively-updated feasible sets. 

\subsection{Gaps and errors in \cite{Fang-16}}

At this point, it is important to note that:

1. (P3) as well as its iterative counterpart (P3a) are \textit{never} strictly convex since their objective $R(\bQ,t)=C_0(\bQ)-t$ is linear in $t$ (see e.g. \cite[p. 67]{Boyd-04} for the definition of strict convexity). Hence, Lemma 1 of \cite{Fang-16} is based on the false assumption of strict convexity of (P3). This also applies to (P4), Lemma 2, and (P7), all in \cite{Fang-16}.

2. The key inequality \cite[eq.(17)]{Fang-16} is claimed to be true due to "the strict convexity of problem (P3)", but no proof is provided for this claim. This also applies to Lemma 2 of \cite{Fang-16}. As our analysis below demonstrates (see Proposition 2), strict convexity is neither necessary nor sufficient for this claim to be true.

3. While \cite[eq.(17)]{Fang-16} in Lemma 1 is using the approximate secrecy rate $R(\bQ^v,\bW^v)$, Algorithm 1 computes the true secrecy rate $C_s(\bQ^v,\bW^v)$ (not its approximation $R$). Hence, even if the analysis of Lemma 1 were correct, it would not imply that Algorithm 1 converges, since \cite{Fang-16} never proves that the sequence $\{C_s(\bQ^v,\bW^v)\}$ converges.

4. In Lemma 1, it is claimed, without proof, that "the limit point of sequence $(\bQ^v,\bW^v)$ constitutes a maxima of problem (P2)".
To see that this claim is not justified, observe that while the approximate problem (P3) is convex, the original problem (P2) is not (unless all $\bH_k=0$), since $C_k(\bQ,\bW)$ are not convex\footnote{We remark that if this claim were correct, it would imply that any non-convex problem can be efficiently solved via its successive convex approximation. This is clearly not the case, see e.g. \cite{Horst-95}\cite{Tuy-16}.}. To illustrate the difficulties of non-convex optimization based on local convex approximations and what may go wrong in the process, let us consider the following simple (scalar) problem:
\bal
\label{eq.pfi}
\max_{a\le x \le b} f(x),\ f(x) = \max_i\{f_i(x)\}, \ i=1...4,
\eal
where each $f_i(x)$ is concave but $f(x)$ is not and hence this problem is not convex, as illustrated in Fig. 1. This problem is equivalent to
\bal
\label{eq.pf}
\max_{t,x} t,\ \mbox{s.t.}\ t \le f(x),\ a\le x \le b,
\eal
which is also not convex, since $f(x)$ is not concave. Since $f(x)$ is concave on each sub-interval (see Fig. 1), the SCA algorithm, when applied to either \eqref{eq.pfi} or \eqref{eq.pf}, converges in just one step, but its convergence point $x_c$ and the respective objective value $f(x_c)$ depend on a starting point $x_0$. If $x_0=x_{01}$, then $x_c=x_{01}$ and $f(x_c)=y_1$, which is the global \textit{minimum} (not maximum). Likewise, if $x_0=x_{04}$, then $x_c=x_{04}$ and $f(x_c)=y_4$, which is a local minimum (not maximum). If $x_0=x_{02}$, then $f(x_c)=y_2$, which is a local maximum, and if $x_0=x_{03}$, then $f(x_c)=y_3$, which is the global maximum. Hence, a convergence point and its respective objective value depend significantly on the initial point when the original problem is not convex. Without  further assumptions and analysis (missing in \cite{Fang-16}), no claims can be made about the relationship of local and global optima, which is well-known in the optimization literature \cite{Boyd-04}-\cite{Tuy-16}.

\begin{figure}[htbp]
\centerline{\includegraphics[width=3.5in]{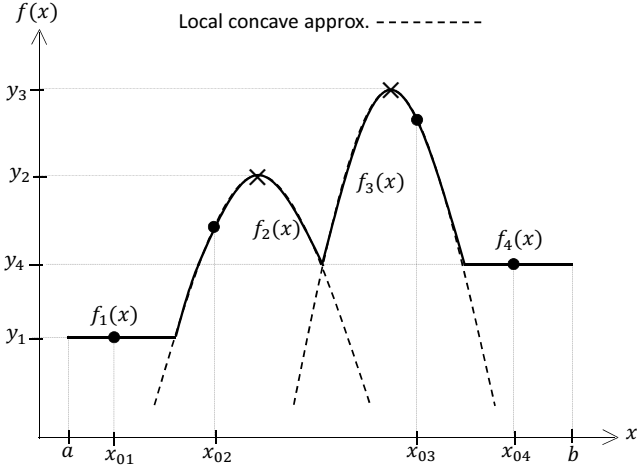}}
\caption{The difficulties of successive concave approximation for a non-concave objective. $f(x)$ is solid bold; $f_2(x)$ and $f_3(x)$ are dashed bold.}
\label{fig.1}
\end{figure}

To rigorously establish this claim of Lemma 1, one has to demonstrate that (i) a limit point of the sequence $\{\bQ^v,\bW^v\}$ generated by the approximate problem (P3a) solves the KKT conditions of the original problem (P2), and that (ii) the KKT conditions of (P2) are sufficient for optimality. Since (P2) is never convex (unless all $\bH_k=0$ -- a trivial case not considered here), item (ii) is out of reach (see e.g. \cite{Boyd-04}-\cite{Tuy-16}). We overcome this difficulty by reformulating (P2) and adopting additional assumptions, as explained in the next section.

5. Finally, a termination criterion for Algorithms 1 and 2 is missing in \cite{Fang-16}. This is important since the convergence of an algorithm and its proof significantly depend on its termination criterion (an algorithm may converge under one criterion and not converge under another). A suitable termination criterion is given in \eqref{eq.A1.stop} below.

These gaps and errors are corrected below.

\section{Corrections}

We will need below the following technical results related to (P2), (P3) and (P3a).

\begin{prop}
\label{Prop.1}
The following holds:
\bal
\label{eq.Prop.1.1}
&(\bZ,t) \in S_{Z,t}(\bZ^v) \,\,  \Rightarrow \,\,  \bZ \in S_{Z}\ \forall v \\
\label{eq.Prop.1.2}
&C_{lk}(\bZ|\bZ)=C_k(\bZ) \,\, \forall \,\, \bZ\\
\label{eq.Prop.1.3}
&0 \le C_k(\bZ)  \le  C_{lk}(\bZ|\widetilde{\bZ}) \,\, \forall \,\, \bZ,\widetilde{\bZ}\\
\label{eq.Prop.1.4}
&t^v = \max_{k \in \mathcal{K}}C_{lk}(\bZ^v|\bZ^{v-1})
\eal
\begin{proof}
Note, from \eqref{eq.S_Q,W} and \eqref{eq.S_Q,W,t|Q^v,W^v}, that $S_{Z,t}(\bZ^v)$ contains all the constraints of $S_{Z}$, in addition to $C_{lk}(\bZ|\bZ^v) \le t$. This implies \eqref{eq.Prop.1.1}.

To show \eqref{eq.Prop.1.2}, observe that
\bal
C_{lk}(\bZ|\bZ)&=\overline\varphi_k(\bZ|\bZ)-\phi_k(\bW) \notag \\ \notag
&=\varphi_k(\bZ) +tr(\bT_k(\bZ)(\bQ-\bQ))\\ \notag
&\quad + tr(\bR_k(\bZ)(\bW-\bW)) -\phi_k(\bW) \notag \\
&=\varphi_k(\bZ)-\phi_k(\bW) =C_k(\bZ)
\eal

To show \eqref{eq.Prop.1.3}, use the following argument:
\bal
0 &\stackrel{\text{(a)}}{\le}  \varphi_k(\bZ)-\phi_k(\bW) =C_k(\bZ)  \notag \\
& \stackrel{\text{(b)}}{\le} \overline\varphi_k(\bZ|\widetilde{\bZ})-\phi_k(\bW)
=C_{lk}(\bZ|\widetilde{\bZ})
\eal
where (a) follows from \eqref{eq.varphi_k}-\eqref{eq.phi_k} and the fact that $\ln|\bI+\bX|$ is increasing in $\bX \ge 0$ (see e.g. \cite{Zhang-99}), so that $\varphi_k(\bZ) \ge \phi_k(\bW)$;
(b) follows since $\varphi_k(\bZ)$ is a differentiable, concave function and hence is upper-bounded by its first-order Taylor expansion \cite{Boyd-04}, so that $\varphi_k(\bZ) \le \overline\varphi_k(\bZ|\widetilde{\bZ})$.

\eqref{eq.Prop.1.4} follows from the constraint $t\ge C_{lk}(\bZ|\bZ^{v-1})$ of (P3), (P3a), and the fact that $R(\bQ,t)$ is strictly decreasing in $t$.
\end{proof}
\end{prop}

Now, we are in a position to prove rigorously that Algorithm 1 in \cite{Fang-16} generates an increasing and bounded sequence of true secrecy rates $\{C_s(\bZ^v)\}$ (not just their concave approximations) and hence converges.
To do so, we adopt the following intuitive stopping criterion:
\bal
\label{eq.A1.stop}
\Delta C_{v,v_0}=C_s(\bZ^{v})-C_s(\bZ^{v-v_0}) \le \epsilon
\eal
with some $\epsilon >0$ and $v_0 \ge 1$, so that the algorithm stops at step $v \ge v_0$ if there is no significant improvement over the last $v_0$ steps, where $\epsilon$ is the desired tolerance (accuracy) level. Note that this criterion is using the true secrecy rate $C_s(\bZ^{v})$,  not its concave approximation $R(\bQ^v,t^v)$, which is consistent with Algorithm 1 in \cite{Fang-16}. Further note that if the sequence $\{C_s(\bZ^v)\}$ converges, then this criterion will eventually be satisfied.

\begin{prop}
\label{prop.2}
Let $(\bZ^{v}, t^{v})$ be computed iteratively according to problem (P3a) in \eqref{eq.Q^v,W^v,t^v}. Then, (i) both the true secrecy rate $ C_s(\bZ^v)$ and its concave approximation $R(\bQ^v,t^v)$ are increasing and upper bounded by $C_s^*$ and, hence, converge; (ii) Algorithm 1 in \cite{Fang-16} also converges under the stopping criterion in \eqref{eq.A1.stop}, for any $\varepsilon >0$ and any $v_0 \ge 1$.

\begin{proof}
To prove 1st claim, note the following:
\bal
C_s^* &=\max_{\bZ \in S_{Z}} C_s(\bZ) \\
\label{eq.P2.2}
&\ge C_s(\bZ^v) \\
\label{eq.P2.3}
&=C_0(\bQ^v)-\max_{k \in \mathcal{K}}C_k(\bZ^v) \\
\label{eq.P2.4}
&\ge C_0(\bQ^v)-\max_{k \in \mathcal{K}}C_{lk}(\bZ^v|\bZ^{v-1}) \\
\label{eq.P2.5}
&=  C_0(\bQ^v)- t^v =R(\bQ^v,t^v)\\
\label{eq.P2.6}
&=\max_{(\bZ,t) \in S_{Z,t}(\bZ^{v-1})} (C_0(\bQ)-t)\\
\label{eq.P2.7}
&= \max_{\bZ \in S_{Z}}  \big(C_0(\bQ)- \max_{k \in \mathcal{K}}C_{lk}(\bZ|\bZ^{v-1})\big)\\
\label{eq.P2.8}
&\ge C_0(\bQ^{v-1})- \max_{k \in \mathcal{K}}C_{lk}(\bZ^{v-1}|\bZ^{v-1}) \\
\label{eq.P2.9}
&= C_0(\bQ^{v-1})- \max_{k \in \mathcal{K}}C_k(\bZ^{v-1}) \\
\label{eq.P2.10}
&= C_s(\bZ^{v-1}) \\
\label{eq.P2.11}
&\ge C_0(\bQ^{v-1})- t^{v-1} =R(\bQ^{v-1},t^{v-1})
\eal
where \eqref{eq.P2.2} follows from $\bZ^v \in S_Z$ (due to \eqref{eq.Prop.1.1});
\eqref{eq.P2.4} is due to \eqref{eq.Prop.1.3};
\eqref{eq.P2.5} follows from \eqref{eq.Prop.1.4};
\eqref{eq.P2.7} follows from the fact that this problem is equivalent to \eqref{eq.P2.6} by eliminating slack variable $t$;
\eqref{eq.P2.8} is due to $\bZ^{v-1} \in S_{Z}$, see \eqref{eq.Prop.1.1};
\eqref{eq.P2.9} is due to \eqref{eq.Prop.1.2};
\eqref{eq.P2.11} follows from \eqref{eq.P2.3}-\eqref{eq.P2.5} via $v \to v-1$.

2nd claim follows from 1st one, since every convergent sequence is a Cauchy sequence, see e.g. \cite{Bronshtein-86}.
\end{proof}
\end{prop}

It should be pointed out that the proof of this Proposition is completely different from the (incorrect) "proof" of Lemma 1 in \cite{Fang-16}. Strict convexity is neither necessary nor sufficient for this proof.

The following interlacing property of the sequence generated by (P3a) (or, equivalently, by Algorithm 1 in \cite{Fang-16}) follows from the proof of Proposition 2:
\bal\notag
C_s^* &\ge C_s(\bZ^v) \ge R(\bQ^v,t^v)\\
 &\ge C_s(\bZ^{v-1}) \ge R(\bQ^{v-1},t^{v-1})\ge 0\ \ \forall v \ge 1
\eal
from which it is clear that both the true secrecy rates $C_s(\bZ^v)$ and their concave approximations $R(\bQ^v,t^v)$ form increasing, bounded sequences and hence converge.

We caution  the reader not to conclude from Proposition 2 that a convergence point of (P3a) also solves (P2), i.e. that it is the global or even a local maximum of (P2). Indeed, since (P3a) is just a convex approximation of non-convex problem (P2), its convergence point $\bZ_c$

(i) depends on initial (staring) point $\bZ^0$, $\bZ_c = \bZ_c(\bZ^0)$, and so is the achieved secrecy rate $C_s(\bZ_c(\bZ^0))$ at that point (i.e. using different initial points $\bZ^0$ may result at different achieved rates and there is no guarantee that any of them is the global maximum of (P2) or even close to it\footnote{This is a well-known general property of all sequential algorithms applied to non-convex problems, see. e.g. \cite{Boyd-04}-\cite{Tuy-16}. This property is also illustrated in Fig. 1. In general, for non-convex problems, algorithms with guaranteed convergence to global optima are of exponential complexity \cite{Horst-95}.}), and

(ii) a convergence point of (P3a) can be a local or even global minimum (not maximum) of (P2), see Fig. 1 for an illustration of this phenomenon.

However, under certain additional assumptions on (P1), one can eliminate these unpleasant possibilities and establish the relationship between a convergence point of (P3a) and the global maximum of (P1) and (P2), as shown below.

To establish this relationship, let us modify (P2) by re-defining its slack variable as follows: $t_2=C_0(\bQ)-t$, so that
\bal
(P2a)\ \ C_s^*=&\max_{(\bZ,t_2) \in S_{Z,t_2}} t_2
\eal
where $t_2$ is the new slack variable and the new feasible set is
\bal\notag
\label{eq.S_Q,W,t_2}
S_{Z,t_2}=\big\{&(\bZ,t_2):\ \bZ \in S_Z,\\
 &t_2- C_0(\bQ) +C_k(\bZ) \le 0 \ \forall k \in \mathcal{K}\}
\eal
Using the same approach as in Proposition 1 (see \eqref{eq.Prop.1.4}), it is not difficult to show that, at an optimal point of (P2a), $t_2^* = \min_{k \in \mathcal{K}} \big(C_0(\bQ^*)-C_k(\bZ^*)\big)$, so that problem (P2a) is indeed equivalent to (P1) and $t_2^*$ is the maximum achievable secrecy rate.
The next proposition shows that a convergence point of (P3a) is a KKT point of (P2a) (i.e. it solves the KKT conditions of (P2a)) and hence a stationary point of (P1).

\begin{prop}
\label{prop.KKT.P3.P2a.P1}
Let $\bZ_c$ be a convergence point of problem (P3a), that is
\bal
\bZ_c \triangleq \bZ^v=\bZ^{v-1}
\eal
for some $v$ (i.e. the iterative procedure of (P3a) converges at iteration $v-1$). Then, $\bZ_c$ is also a KKT point of problem (P2a).
\begin{proof}
The proof is by examining the KKT conditions of both problems and showing that they coincide at convergence point $\bZ=\bZ_c$. Hence, $\bZ_c$ also solves the KKT conditions of (P2a), which is equivalent to (P2) and (P1). See Appendix for details.
\end{proof}
\end{prop}

At this point, we remark that this Proposition does not imply that a convergence point of (P3a) also solves (P2a), i.e. is its global maximum, since (P2a) is not a convex problem in general and therefore its KKT conditions are not sufficient for global optimality \cite{Boyd-04}. However, under certain additional assumptions, (P2a) is a convex problem and hence such implication does hold, as the next proposition shows.

\begin{prop}
\label{prop.converge.opt}
Let $C_0(\bQ)-C_k(\bZ)$ be concave functions of $\bZ=(\bQ,\bW)$ for all $k \in \mathcal{K}$. Then, any convergence point of (P3a) is also globally optimum for (P2a) and hence for the original problem (P1).
\begin{proof}
If all $C_0(\bQ)-C_k(\bZ)$ are concave functions, then (P2a) is a convex problem, since all inequality constraint functions are convex and the objective is a concave function. In this case, its KKT conditions are sufficient for global optimality \cite{Boyd-04}. Using Proposition \ref{prop.KKT.P3.P2a.P1}, any convergence point of (P3a) is globally-optimal for (P2a) and, hence, for equivalent problems (P2) and (P1).

At this point, we remark that the same conclusion cannot be obtained with (P2), since it is never a convex problem (except for the trivial case of $\bH_k=0\ \forall k$), even under the stated conditions when (P2a) is convex.
\end{proof}
\end{prop}

The functions $C_0(\bQ)-C_k(\bZ)$ are concave for some class of degraded wiretap channels, see e.g. \cite{Khisti-10}-\cite{Loyka-16-2}. Further note that, under the conditions of Proposition 4, the original problem (P1) is also convex, since its objective is concave as a point-wise minimum of concave functions,
\bal\notag
C_s(\bZ) = C_0(\bQ)-\max_{k \in \mathcal{K}}C_k(\bZ) = \min_{k \in \mathcal{K}}(C_0(\bQ)-C_k(\bZ))
\eal
but, at the same time, (P2) is not a convex problem.

It should be noted that if $C_0(\bQ)-C_k(\bZ)$ is not concave for some $k$, then a convergence point of (P3a) is not necessarily globally-optimum  for (P2a) and (P1), since the KKT conditions of (P2a) are not sufficient for global optimality in this case. 

For the special cases of (i) a single eavesdropper, or (ii) multiple eavesdroppers when there exists a dominant one, or (iii) when they cooperate, algorithms with guaranteed convergence to the global optimum (even if (P2a) is not convex) were presented in \cite{Dong-20}. The general case of multiple non-cooperating eavesdroppers (without dominant one) remains an open problem.

Finally, it is straightforward to see (following the steps of the proofs above) that the conclusions of Propositions 2-4 also hold for problems (P4) and (P7) in \cite{Fang-16}.

\subsection{Appendix: Proof of Proposition \ref{prop.KKT.P3.P2a.P1}}

The Lagrangian of (P2a) is as in \eqref{eq.Lp2a}, where $\mu_1$, $\mu_2$, $\lambda_k$, $\bM$ and $\bN$ are the Lagrange multipliers (dual variables) responsible for the interference power constraint $P_I(\bZ) = tr(\bG\bQ\bG^H+\bG\bV\bW\bV^H\bG^H)\le \Gamma$, the transmit power constraint $tr(\bQ+\bW) \le P$, the slack variable constraint $t_2-\big(C_0(\bQ)-C_k(\bZ)\big) \le 0$, and the positive semi-definite constraints $\bQ, \bW \succeq 0$, respectively. The respective KKT conditions are in \eqref{cond2.kkt1}-\eqref{cond2.kkt5} (these conditions are similar to those in \cite{Khisti-10}-\cite{Loyka-16-2} but also account for the difference in the problem statement here with the extra constraints and variables).
\begin{figure*}
\bal
\label{eq.Lp2a}
&L_{P2a} = -t_2 +\mu_1\big(P_I(\bZ)-\Gamma\big) +\mu_2(tr(\bQ+\bW)-P)- tr(\bM\bQ+\bN\bW) +\sum_{k \in \mathcal{K}}\lambda_k\big(t_2+C_k(\bZ)-C_0(\bQ)\big)\\
\label{cond2.kkt1}
&\nabla_{\bQ} L_{P2a} =\mu_1\bG^H\bG+\mu_2\bI - \bM +\sum_{k \in \mathcal{K}}\lambda_k \big(\nabla_{\bQ}C_{k}(\bZ)-\nabla_{\bQ}C_0(\bQ)\big)=0\\
\label{cond2.kkt2}
&\nabla_{\bW} L_{P2a} = \mu_1\bV^H\bG^H\bG\bV+\mu_2\bI - \bN+\sum_{k \in \mathcal{K}}\lambda_k\nabla_{\bW}C_{k}(\bZ)=0,\\
\label{cond2.kkt3}
&\nabla_{t} L_{P2a}=-1+ \sum_{k \in \mathcal{K}}\lambda_k=0\\
\label{cond2.kkt4}
&\mu_1\big(P_I(\bZ)-\Gamma\big)=0, \ \mu_2(tr(\bQ+\bW)-P)=0,\ \bM\bQ=0, \ \bN\bW=0,\ \lambda_k\big(t_2+C_k(\bZ)-C_0(\bQ)\big)=0 \ \forall k \in \mathcal{K}\\
\label{cond2.kkt5}
&\mu_1,\mu_2 \ge 0, \ \bM, \bN \succeq 0,\ \bQ, \bW \succeq 0,\ P_I(\bZ) \le \Gamma,\ tr(\bQ+\bW) \le P,\ \lambda_k \ge 0,\ t_2\le C_0(\bQ)-C_k(\bZ)\ \forall k \in \mathcal{K}
\eal
\hrule
\end{figure*}

Next, the Lagrangian $L_{P3a}$ of (P3a) or (P3) is as in \eqref{eq.L} and
the respective KKT conditions are in \eqref{cond.kkt1}-\eqref{cond.kkt5}, where we used the rules of matrix differentiation with respect to $\bQ, \bW$, see e.g. \cite{Boyd-04}\cite{Zhang-99}\cite{Khisti-10}-\cite{Loyka-16-2}. Here, \eqref{cond.kkt1}-\eqref{cond.kkt3} are the stationary conditions, \eqref{cond.kkt4} and \eqref{cond.kkt5} are the complementary slackness conditions and the primal/dual feasibility conditions. Note that \eqref{cond.kkt3} implies that $\lambda_k > 0$ for at least one $k \in \mathcal{K} $. Using this fact, the constraints $C_{lk}(\bZ|\bZ^{v-1})\le t$ and $\lambda_k\big(C_{lk}(\bZ|\bZ^{v-1})-t\big)=0$, we observe that $t^v=\max_{k \in \mathcal{K}}C_{lk}(\bZ|\bZ^{v-1})$, which is in agreement  with \eqref{eq.Prop.1.4}.

\begin{figure*}
\bal
\label{eq.L}
&L_{P3a} = -C_0(\bQ) +t+\mu_1\big(P_I(\bZ)-\Gamma\big)+\mu_2(tr(\bQ+\bW)-P) - tr(\bM\bQ+\bN\bW)+\sum_{k \in \mathcal{K}}\lambda_k\big(C_{lk}(\bZ|\bZ^{v-1})-t\big)\\
\label{cond.kkt1}
&\nabla_{\bQ} L_{P3a} = -\nabla_{\bQ} C_0(\bQ)+ \mu_1\bG^H\bG+\mu_2\bI - \bM+\sum_{k \in \mathcal{K}}\lambda_k\nabla_{\bQ}C_{lk}(\bZ|\bZ^{v-1})=0\\
\label{cond.kkt2}
&\nabla_{\bW} L_{P3a} = \mu_1\bV^H\bG^H\bG\bV+\mu_2\bI - \bN+\sum_{k \in \mathcal{K}}\lambda_k\nabla_{\bW}C_{lk}(\bZ|\bZ^{v-1})=0\\
\label{cond.kkt3}
&\nabla_{t} L_{P3a}=1- \sum_{k \in \mathcal{K}}\lambda_k=0\\
\label{cond.kkt4}
&\mu_1\big(P_I(\bZ)-\Gamma\big)=0, \ \mu_2(tr(\bQ+\bW)-P)=0,\ \bM\bQ=0, \ \bN\bW=0,\ \lambda_k\big(C_{lk}(\bZ|\bZ^{v-1})-t\big)=0\ \forall k \in \mathcal{K}\\
\label{cond.kkt5}
&\mu_1,\mu_2 \ge 0, \ \bM, \bN \succeq 0,\ \bQ, \bW \succeq 0,\ P_I(\bZ) \le \Gamma,\ tr(\bQ+\bW) \le P,\ \lambda_k \ge 0, \ C_{lk}(\bZ|\bZ^{v-1})\le t \ \forall k \in \mathcal{K}
\eal
\hrule
\end{figure*}

Now observe that, at a convergence point $\bZ=\bZ^v=\bZ^{v-1}$ of (P3a),
\bal
\nabla_{\bQ} C_{lk}(\bZ|\bZ^{v-1}) &=\bT_k(\bZ^{v-1})= \bT_k(\bZ) =\nabla_{\bQ} C_k(\bZ) \\
\label{eq.3rd.cond.stationary}
\nabla_{\bQ}C_0(\bQ)&=\sum_{k \in \mathcal{K}}\lambda_k \nabla_{\bQ} C_0(\bQ)
\eal
where the last equality is due to \eqref{cond.kkt3}. Hence, $\nabla_{\bQ} L_{P3a} = \nabla_{\bQ} L_{P2a}$.
Next, observe that
\bal
\nabla_{\bW} C_{lk}(\bZ|\bZ^{v-1}) &=\bR_k(\bZ^{v-1})-\nabla_{\bW} \phi_k(\bW) \notag \\
&=  \bR_k(\bZ)-\nabla_{\bW} \phi_k(\bW) = \nabla_{\bW} C_k(\bZ)
\eal
Hence, $\nabla_{\bW} L_{P3a} = \nabla_{\bW} L_{P2a}$. Therefore, the stationarity conditions \eqref{cond2.kkt1}-\eqref{cond2.kkt3} and \eqref{cond.kkt1}-\eqref{cond.kkt3} of both problems are the same. The rest of the conditions also coincide at a convergence point, since $C_{lk}(\bZ|\bZ)=C_k(\bZ)$ (see \eqref{eq.Prop.1.2}) and $t_2=C_0(\bQ)-t$. Hence, a solution of the KKT conditions of (P3a) at its convergence point also solves the KKT conditions of (P2a), which is equivalent to (P2) and (P1).


\end{document}